\documentstyle[aps,epsfig,amstex,psfrag]{revtex}
\newcommand{\be}{ \begin{equation} }
\newcommand{\ee}{ \end{equation} }
\newcommand{\bea}{ \begin{eqnarray} }
\newcommand{\eea}{ \end{eqnarray} }
\newcommand{\mel}[3]{\langle #1 | #2 | #3 \rangle } 
\newcommand{\ket}[1]{| #1 \rangle } 
\newcommand{\bra}[1]{\langle #1 | } 
\newcommand{\spinup}{|\!\uparrow\rangle } 
\newcommand{\spindown}{|\!\downarrow\rangle }

\newcommand{\bfrho}{\rho} 
\newcommand{\tr}{\mathop{\rm tr}\nolimits}
\begin{document}
\title{Decoherence and dissipation during a quantum XOR gate operation}
\author{Michael Thorwart$^{1,2}$ and Peter H\"anggi$^{1}$}
\address{
              $^1$ Institut f\"ur Physik -
              Universit\"at  Augsburg, Universit\"atsstr.\ 1, 86135 Augsburg,
              Germany \\
	      $^2$ Department of Applied Physics, 
	      Delft University of Technology, Lorentzweg 1, 2628 CJ Delft,
              The Netherlands \\
}
\date{Date: \today}
\maketitle
\begin{abstract}
The dynamics of a quantum XOR gate operation in a two-qubit 
system being coupled to a bath of quantum harmonic oscillators is 
investigated. Upon 
applying the numerical quasiadiabatic 
propagator path integral method, we obtain the numerically precise 
 time-resolved evolution of this interacting two-qubit system in presence of 
time-dependent external fields without further 
approximations. We simulate  the dissipative gate operation for characteristic  
experimental realizations  of condensed matter qubits; namely, the 
flux and charge qubits realized in superconducting Josephson systems and 
qubits formed with semiconductor quantum dots. 
Moreover, we study systematically the quality of the  XOR gate  
by determining the four characteristic gate quantifiers: fidelity,  
purity, 
the quantum degree, and 
the entanglement capability 
of the gate. Two different types 
of errors in the qubits have been modelled, i.e., bit-flip errors 
and phase errors. The dependence of the quality of the gate operation 
on the environmental {\em temperature\/}, on the {\em friction strength\/} 
stemming from the 
system-bath interaction, 
and on the strength of the {\em interqubit coupling\/} 
is systematically explored: Our main finding is that the four gate quantifiers 
depend only weakly on temperature, but are rather sensitive to the friction 
strength.  \\[1mm]
PACS:  03.67.Lx, 03.65.Yz, 89.70.+c, 05.30.-d,  05.40.-a
%

\end{abstract}
\vspace{-2mm}
\pacs{PACS: 03.67.Lx, 03.65.Yz, 89.70.+c, 05.30.-d,  05.40.-a}
\section{Introduction}
%


The basic elements of quantum computation  
are logic quantum gates which represent manipulations of quantum bits,  
$\ket{0}$ and $\ket{1}$, 
according to Boolean algebra. Any arbitrary 
complex  logic operation can be build up of only a few basic gates 
({\em universal gates\/}) \cite{Deutsch89} and one can show that almost 
every gate which operates on two or more qubits is a universal gate 
\cite{Lloyd95}. The explicit construction of quantum networks for elementary 
arithmetic operations 
then becomes possible upon appropriately combining such universal gates;  
see, for instance, Ref.\ \cite{Vedral96} for the explicit construction of 
 the addition or the modular exponentiation. 
In turn, this permits the implementation of 
Shor's quantum factorizing algorithm \cite{Shor94} in terms of elementary gates. 
Together with Deutsch's algorithm \cite{Deutsch85}, 
these two quantum algorithms are presently 
the most important examples  which are  
known to be superior to their classical counterparts and which do  
justify the current efforts towards a technological realization of a quantum 
computer.  

In this work we concentrate on 
one such elementary gate, namely the {\em quantum exclusive OR (XOR) gate\/}. 
It is a unitary transformation 
which propagates an initial state $|\Psi_{\rm in}\rangle$ 
of a two-qubit system  to a final state 
$|\Psi_{\rm out}\rangle = U_{\rm XOR} |\Psi_{\rm in}\rangle$. 
Represented in the computational basis 
$|b_i\rangle \in \{|00\rangle, |01\rangle, |10\rangle, 
|11\rangle\}$ ($i=1,...,4$), the XOR gate operation can be written as
\begin{equation}
U_{\rm XOR} = \left( 
\begin{array}{cccc}
1 & 0 & 0 & 0 \\
0 & 1 & 0 & 0 \\
0 & 0 & 0 & 1 \\
0 & 0 & 1 & 0 
\end{array}
\right) \, . 
\label{uxor}
\end{equation}
%
Since this operation inverts the state of the second qubit of the basis states 
if the first qubit is in the state $\ket{1}$, this operation 
is also called the {\em quantum controlled NOT (CNOT) gate\/}. 
The set of all one-qubit gates together with the quantum 
XOR gate is universal, as has been demonstrated in Ref.\ 
\cite{Barenco95}. 

The main impediment on the roadway to a working quantum computer is decoherence 
\cite{Unruh95,Garg96,Palma96,Landauer98,Loss98a,Makhlin00}. 
It disturbs the phase relation in a quantum superposition state 
 and therefore is effective at the roots where the quantum computer 
is believed to have its most important advantage. 
Any realistic quantum computer 
will have some interaction with its environment which induces 
decoherence (decay of the off-diagonal elements of the reduced 
density matrix) and dissipation (change of populations of the reduced 
density matrix). 
Moreover, other sources for decoherence which are due to imperfect 
gate operations and  to cross-talks of the qubits within a register 
need to be considered \cite{Landauer98}. 

Several previous works in the literature 
deal with the effect of decoherence in quantum information processing systems. 
Unruh \cite{Unruh95} and Palma {\em et al.\/} \cite{Palma96} consider a 
 model of a single qubit which is represented by the eigenstates of the  
quasi-spin operator $\sigma_z$ and which couples  to a bosonic environment via its 
$\sigma_z$-component. It describes appropriately the dephasing (decoherence) 
but does 
not include population exchange (dissipation). Combining $L$ {\em non-interacting\/}  
qubits of this type, they estimate the decoherence (in the limit of a large 
coherence length of the bath) to increase exponentially with the length $L$ 
of the register. 

Dissipative effects (bit flip errors) are properly described 
by the so-called spin-boson model 
\cite{Leggett,Weiss99,Grifoni98,Grifoni99}, 
where the qubit is represented by the 
$\sigma_x$-component of the spin-1/2, but the coupling to the bosonic bath 
is mediated by the $\sigma_z$-component of the spin-1/2 
(note, that this refers to the localized representation). In this model, 
the bath also induces transitions between the two system eigenstates 
(bit flips) and -- in addition to decoherence --   
 energy is exchanged between system and bath. The general 
solution of the problem in terms of a generalized  (non-Markovian) master 
equation for the entire reduced density matrix for an 
arbitrary initial preparation in presence of a static bias and also for 
a time-dependent driving has been given in Ref.\  \cite{Grifoni99}. 
In our work, 
the assumption of a Markovian bath and of a weak system-bath interaction 
(Bloch-Redfield approach), 
 which may restrict the validity of the master equation 
 (see below), is not made.

The previously discussed works concern the investigation of 
decoherence in single qubits or in a register 
of non-interacting qubits.  Decoherence and dissipation in a system of 
{\em interacting\/} qubits has been studied only rarely. 
The dynamics of two coupled two-level systems 
  has been investigated by Dub\'{e} and Stamp 
 \cite{Stamp98} by means of a general model for coupled Josephson junctions, for 
 coupled nanomagnets or for interacting Kondo impurities. 
 Each two-level system is represented (in the tunneling 
 representation) by the  $\sigma_x$-component of a spin-1/2. 
 The two spins interact 
 via their $\sigma_z$-components. Moreover, their $\sigma_z$-components 
 couple to a bosonic bath. By use of real-time path integrals the dynamics 
 of the relaxation process is determined. 
  Although no specific problem of quantum information 
 processing is investigated,  this is the first work where 
 two interacting spins 
 in a dissipative bath have been considered. 
 
 A similar model has been studied 
 by Governale, Grifoni and Sch\"on \cite{Governale01}. 
 Two biased spin-1/2 systems interact via their $\sigma_y$-components 
 which is the 
 appropriate coupling for Josephson junction charge qubits (see below).  
 Moreover, their $\sigma_z$-components couple either to 
 the same or to different 
 bosonic baths. Applying the widely used 
 Bloch-Redfield formalism, the time evolution of the populations of the logical 
 states is evaluated. This model describes dissipation being caused by fluctuations 
 in  voltage sources in Josephson junction charge qubits (see below). However, 
 no specific quantum information operation has been considered. 
 
 A two-qubit quantum gate for quantum information processing in coupled quantum 
 dots has been investigated in Refs.\  \cite{Loss98a,Ahn00}. 
 Two  spin-1/2 systems are coupled using a time-dependent 
 Heisenberg-type interaction. Moreover, a coupling of the spins to a bosonic 
 bath has been taken into account. By solving the quantum Liouville equation 
 in the limit of weak system-bath coupling (Born-Markov approximation) for  
 the reduced density operator, the purity and the fidelity 
 of the swap operation $U_{\rm swap} \ket{ij} = \ket{ji} (i,j=0,1)$ 
 is calculated   
 as a function of time. However, the authors consider 
 the time-evolution of the quantum system 
 {\em after the swap operation has been completed\/}. The same is true for the 
 XOR gate operation in Ref.\  \cite{Loss98a}, where, additionally, a further 
 assumption has been made: The pulse sequence to realize the quantum XOR consists of 
 four pulses of the external fields. Each pulse is taken to be constant over the 
 corresponding time interval. To obtain the solution over the entire time span 
 within the Born-Markov approximation, 
 it is necessary to assume a finite time interval between the single pulses. 
 This is required because the Born-Markov approximation is known to violate 
 positivity of the reduced density operator at short transient 
 times \cite{vanKampen92,Suarez92}.  
 This additional time span (pulse-to-pulse time) 
 has been taken as three times the switching time interval. 
 This leads to an extension of the computation time which is only  
 due to formal mathematical reasons and which deteriorates the quality of the 
 gate operation. Moreover, 
 a systematic study of the dependence of the gate quantifiers on the 
 relevant parameters has not been given.


In this work, we investigate systematically the XOR quantum gate in presence 
of an interaction of the qubits with their environment. Thereby, we take 
into account the full time-dependence of the external fields which induce the 
XOR operation without invoking further approximations on the system Hamiltonian. 
In particular, we use the numerical {\em ab-initio\/} technique of 
the quasiadiabtic propagator path integral (QUAPI) \cite{quapi} (for other  
applications, see also Refs.\ \cite{ThorwartQUAPI,ThorwartQUAPI2}). 
This numerically precise iterative 
real-time path integral method   
does not suffer from the above mentioned problem of lacking positivity. 
In order to realize the logic XOR operation in physical systems, 
we introduce a generic model Hamiltonian which is suitable for 
studying the XOR operation on a very general and idealized level. We determine the 
quality of the gate by calculating the four characteristic gate quantifiers 
introduced by Poyatos, Cirac and Zoller \cite{Poyatos97}: namely, the 
(i) purity, 
(ii) fidelity, (iii) quantum degree, and (iv) entanglement capability. 
To that end, we consider two important types of computational errors, i.e., 
phase errors and bit flip errors. The former can be modelled by coupling the 
$\sigma_z$-component of each spin to the bath while 
the later is induced by coupling the $\sigma_x$-component of 
each spin to the bath. We are mainly interested in the quality of the gate 
operation during its time evolution and, most importantly, 
right after it has been completed. Our 
analysis covers a wide field of physical systems and we choose parameter sets 
which mimic the realistic physical situation for flux qubits or charge qubits 
proposed in superconducting hybrid  
systems as well as charge and spin states in coupled 
semiconductor quantum dots (see below). 

So far, we have discussed theoretical aspects of quantum information 
processing. However, those refined and highly elaborate concepts face the question 
of how they can be implemented in experimental hardware. Several proposals to build 
a quantum information processor exist. Prominent candidates are, for instance, 
atoms in optical cavities, ions in linear or Paul traps interacting 
with  laser beams, or nuclear spins in an NMR liquid \cite{qivBuch}. 
Although the experimental techniques in those fields of research are currently 
most advanved, the problem of upscaling of a quantum computer can
seemingly only be 
solved within condensed matter systems which can be embeded in an electronic 
circuit. Promiment systems for condensed matter qubits are flux 
states of a superconducting quantum interference device ({\em flux qubits}) 
\cite{Mooij99}
(see also \cite{Makhlin00}), charge states of 
superconducting islands with Josephson junctions ({\em charge qubits\/}) 
\cite{Makhlin00,Averin98}, 
and spin \cite{Loss98a,Loss99a} 
or charge 
\cite{Blick00} 
states in ultrasmall coupled semiconductor quantum 
dots ({\em quantum dot qubits\/}).   
Moreover, several realizations of qubits in nuclear 
\cite{Kane,Privman01} and electronic \cite{Privman01,Udrea00} spins in semiconductor 
nanostructures have been proposed.
In this work, we concentrate on typical experimental situations for flux and charge 
qubits in Josephson junction devices and for charge qubits in  coupled quantum dots. 

The paper is organized as follows: In the subsequent Section 
\ref{sec.I}, 
we introduce a generic model as a starting point for the quantum XOR 
operation including the interaction with the environment. 
 In Section \ref{sec.III}, we present a brief review on the numerical technique 
 of the quasiadiabatic propagator path integral (QUAPI) which we employ 
 in the following.  In order 
 to determine the quality of the decoherent XOR gate, we use four quantifiers 
 which are introduced in Section \ref{sec.IV}. The results and the 
 conclusions are presented in Sec.\  \ref{sec.V} and Sec.\ \ref{sec.VI}, 
 respectively. 

\section{A generic model for the quantum XOR gate} 
\label{sec.I}
\subsection{The coherent XOR operation}
 
The quantum XOR gate is a two-qubit operation which can be modelled by  
 two coupled spin-1/2 systems represented by the Pauli operators 
 $\vec{\sigma}_j 
= (\sigma_j^x, \sigma_j^y, \sigma_j^z)^{\rm T}, j=1,2$. The two logical 
states of each qubit are represented by the two eigenstates of the 
$\sigma_z$-component of each spin, i.e., $\ket{0}_j \equiv \spinup_j$ and 
 $\ket{1}_j \equiv \spindown_j$. 
We assume that the single 
qubit as well as the coupling between the two qubits 
can be controlled by switching on (local) external fields, for 
instance, magnetic fields.  This  system can 
generically be described \cite{Makhlin00} by the Hamiltonian
\begin{equation}
H_{\rm XOR} (t) = -\frac{\hbar}{2}\sum_{j=1}^2 \vec{B}_j  
(t) \vec{\sigma}_j
+ \hbar \sum_{j\ne k} J(t) \, \sigma_j^+ \sigma_k^- \, .
\label{hamiltonid}
\end{equation}
where $\sigma_j^{\pm}= (\sigma_j^{x} + i \sigma_j^{y})/2$. Moreover, 
$\vec{B}_j (t) = (B_j^x(t), 0, B_j^z(t))^{\rm T}, j=1,2$ 
are time-dependent coupling strengths (with the dimension of a frequency) 
arising from local 
time-dependent external fields at the 
site of the spin $j$ in longitudinal ($z$-) or transverse ($x$-) direction. 
In Eq.\  (\ref{hamiltonid}), 
the coupling between the two qubits is assumed to be symmetric; 
furthermore it should be controllable from the outside  
leading to a time-dependent interaction strength $J(t)$. The particular 
form of the interaction in Eq.\  (\ref{hamiltonid}) is only one example.  
We note that this generic model does not account for the particular details 
of a physical realization of qubits in a specific condensed matter 
system. For each individual system, such as  flux qubits or charge 
qubits, the Hamiltonian looks  different in detail. 
In particular, the coupling term between the two qubits takes 
different forms. However, 
the differences to our generic model in Eq.\  (\ref{hamiltonid}) are 
of minor influence only. 
The general physical behavior will be similar such that our 
generic model serves as an archetype.  

\begin{figure}[th]
\begin{center}
\epsfig{figure=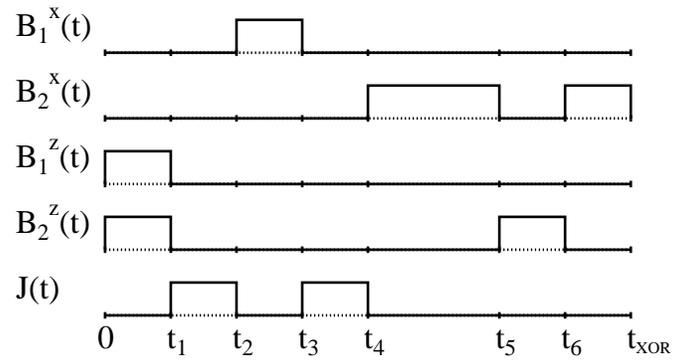,width=90mm,keepaspectratio="true"} 
\end{center}
\caption{Schematic view of a pulse sequence necessary to generate the quantum XOR gate. 
The parameters are set to $B^x=B^z=J={\rm const.}$. 
The frequencies are given in units of $B^z$ while the time is scaled 
in units of $(B^z)^{-1}$. The switching times $t_j$ are given in the text 
and are in this case equal to multiples of $\pi/2$.
 \label{fig.pulse}}
\end{figure}

The quantum XOR gate (\ref{uxor}) can be obtained by a sequence of one- and 
two-qubit operations according to \cite{Makhlin00} 
\begin{eqnarray}
U_{\rm XOR} & = & U_2^x\left(\frac{\pi}{2}\right) U_2^z\left(-\frac{\pi}{2}\right) 
U_2^x\left(-\pi\right) 
U_{12}\left(-\frac{\pi}{2}\right) U_1^x\left(-\frac{\pi}{2}\right) \nonumber \\ 
&  & \times U_{12}\left(\frac{\pi}{2}\right) 
U_1^z\left(-\frac{\pi}{2}\right)U_2^z\left(-\frac{\pi}{2}\right) \, , 
\label{uxorseq}
\end{eqnarray}
where 
\begin{eqnarray}
U_j^{x/z} (\alpha) & = & \exp\left( i \frac{\alpha}{2} \sigma_j^{x/z}\right) \,,  
\qquad j=1,2 \, , \nonumber  \\
U_{12}(\beta) &=& \exp\left(i \beta (\sigma_1^+\sigma_2^- + 
\sigma_1^-\sigma_2^+) \right) 
\label{singprop}
\end{eqnarray}
are the propagators over the single time-intervals with the external fields 
in the Hamiltonian, Eq.\  (\ref{hamiltonid}), switched on and off in the following 
way:  
In order to attain this propagator, a pulse sequence of the external fields 
is necessary. 
For simplicity we assume throughout this 
work, that the pulses are switched instantaneously on and off 
and are constant over the 
time span $t_{\rm off}-t_{\rm on}$ they are on. 
This induces time-dependent interaction strengths  
 ${\cal B}(t)={\cal B} [\Theta( t-t_{\rm on}) - \Theta( t-t_{\rm off})]$ with 
${\cal B}=B_j^{(x/z)}, J = {\rm const.}$ and with $\Theta(t)$ 
being the Heaviside function. 
 Furthermore, we assume that both spins are equal and 
experience local fields of equal strength. This implies 
$B_1^{x/z}=B_2^{x/z}\equiv B^{x/z}$. 
The angles $\alpha$ and $\beta$ in Eq.\  (\ref{singprop}) 
are related to the actual physical propagation time $t$ according to 
\begin{equation}
\alpha=B^{x/z} t  \qquad  \mbox{\rm und} \qquad   \beta = J t  \, . 
\end{equation}
The switching times then follow as 
$
t_1= \pi / (2 B^z), 
t_2=t_1+ \pi / (2 J), 
t_3=t_2+ \pi / (2 B^x), 
t_4=t_3+ \pi / (2 J), 
t_5=t_4+ \pi / (B^x), 
t_6=t_5+ \pi / (2 B^z)$ and 
$
t_{\rm XOR}=t_6+ \pi / (2 B^x), 
$
where $t_{\rm XOR}$ denotes the total time elapsed during the full 
XOR gate operation. 
An example of this pulse sequence is sketched in 
Fig.\  \ref{fig.pulse} for the case of $B^x=B^z=J$. 
The coupling constants are given in units of $B^z$ while the time is scaled 
in units of $(B^z)^{-1}$. One immediately observes that the computation time 
$t_{\rm XOR}$ is extended if the coupling energies are decreased. 
We note that the assumption of rectangular pulses is not required by the 
numerical technique we use and is made here only for the sake of simplicity. 
We could also 
consider other shapes of the pulses which are more realistic for specific 
physical systems, and especially, we could consider imperfect switching 
processes as well; the latter would constitute a further source of decoherence. 
\subsection{Interaction with the environment}
\label{sec.II}
We model the interaction of the qubit system with the fluctuating environment 
by a Hamiltonian, in which $H_{\rm XOR} (t)$ is coupled to a bath of 
harmonic oscillators, i.e., 
\begin{equation}
H(t) = H_{\rm XOR}(t) +  H_{\rm B} + H_{\rm int} ^{x/z} \, ,  
\label{fullham}
\end{equation}
with 
\begin{equation}
H_{\rm B} = \sum_{j=1}^N \hbar \omega_j \left(a^+_j a_j + \frac{1}{2}\right) \, .
\label{bathham}
\end{equation}
Here, $a^+_j (a_j)$ denotes the creation (annihilation) operator of the 
$j$-th bath oscillator with frequency $\omega_j$.  
Since we want to investigate the role of bit-flip errors as well as 
phase errors we include in our model two different types of interactions: 
On the one hand, the $\sigma^{x}$-components of the spins 
 couple to the fluctuating environment and 
the populations of the qubit states are disturbed (bit-flip errors). On the other 
hand, phase errors are generated by a coupling of the $\sigma^{z}$-components of 
the spins to the environmental noise. This is conveniently modelled by the form  
\begin{equation}
H_{\rm int}^{x/z} = \frac{\hbar}{2} \left(\sigma_1^{x/z}+\sigma_2^{x/z}\right) 
 \sum_{j=1}^N \kappa_j^{x/z} (a^+_j + a_j) \, , 
\label{hamint}
\end{equation}
where $\kappa_j^{x/z}$ denotes the coupling strength of the $j$-th oscillator 
to the system and 
where the superscript $(x/z)$ denother {\em either the one or the other kind of 
interaction\/}.
We note that we assume here, to keep things simple, 
 a coupling of the two spins to the same bath. This implies that the spins are 
 effectively coupled to each other via the bath. A coupling of the spins to 
 different (uncorrelated) baths could be readily incorporated in the 
 numerical QUAPI technique (see below). 

To study the dynamics of this system, we have to specify the initial 
conditions.  Throughout this work, we assume that the density operator 
$W(t)$ of the entire 
system-plus-bath at initial time $t=0$ factorizes according to
\begin{equation}
W(0) = \bfrho_{\rm S} (0) \otimes \bfrho_{\rm B} \, .
\label{factorinitial}
\end{equation}
$\bfrho_{\rm S} (0)$ is the density operator of the system at time
$t=0$ and $\bfrho_{\rm B} = Z_{\rm B}^{-1} \exp \left( - 
{H}_{\rm B}/(k_{\rm B} T 
)\right)$ 
is the  canonical equilibrium distribution of the (decoupled) bath  
 at temperature $T$. Moreover, 
  $Z_{\rm B}= \tr \exp (- {H}_{\rm B} /(k_{\rm B} T))$ and 
  $k_{\rm B}$ denotes the Boltzmann constant. 
  
 The influence of the bath is fully characterized  \cite{Weiss99}  
 by the spectral density 
\begin{equation} 
\Gamma^{x/z}(\omega)= 2 \pi \sum_{j=1}^{N}(\kappa_j^{x/z})^2 \delta(\omega-\omega_j)
 \label{discrspede}
\end{equation}
which assumes a continuous form if the number $N$ of oscillators 
 approaches infinity. Throughout this work, we apply an 
Ohmic spectral density with an exponential cut-off, i.e., 
\begin{equation} 
\Gamma^{x/z}(\omega)=\gamma^{x/z} \omega \exp(-\omega/\omega_c) \, ,
 \label{contspede}
\end{equation}
where the dimensionless damping 
parameter $\gamma^{x/z}$ characterizes the strength of 
the interaction with the environment. This spectrum mimics the environmentally 
induced fluctuations in the external circuit which supplies 
 flux through the SQUID loops in the flux qubits 
 \cite{Makhlin00,Mooij99}. 
Moreover, background charge fluctuations in the voltage sources in 
 Josephson charge  qubits 
 \cite{Makhlin00,Averin98} 
 also lead to an Ohmic impedance $R$.  Similarly, 
 electronic states in coupled quantum dot qubits 
 experience an Ohmic environment, 
  either for the spin \cite{Loss98a,Loss99a} 
 or for the charge  \cite{Blick00} 
 degrees of freedom.  
\section{Numerical ab-initio technique: QUAPI}
\label{sec.III}
In order to describe the dynamics of the two-qubit system of 
interest it is sufficient to 
consider the time evolution of the reduced density operator 
\begin{eqnarray} 
\bfrho (t)&= & \tr_{\rm bath} 
 {\cal U}(t,0)\, {W}(0)\, {\cal U}^{-1} (t,0)  \ , \nonumber \\ 
{\cal U}(t,0)&=&{\cal T}\exp \left \{ -i/\hbar \int_{0}^{t} {H}(t') 
 dt' \right \} \  . 
\label{rhored} 
\end{eqnarray} 
Here, ${\cal U}(t,0)$ is the propagator of the full system-plus-bath and 
${\cal T}$ denotes the time-ordering operator. Moreover,  $\tr_{\rm bath}$ 
means the partial trace over the harmonic bath oscillators. Due to our 
assumption 
that the bath is initially at thermal equilibrium and decoupled from 
the system, see Eq.\  (\ref{factorinitial}),  the partial trace over the 
bath can be performed. We denote the matrix elements of the reduced 
density matrix  in the computational basis with 
$\rho_{ij} (t) \equiv \mel{b_i}{\bfrho (t)}{b_j}$ 
and rewrite them according to Feynman and Vernon \cite{Feynman63} as 
\begin{equation} 
\rho_{ij} (t) = \sum_{m,n=1}^4 \ G_{ij,mn}(t,0) \rho_{mn} (0) \ ,
\label{2.1} 
\end{equation} 
with the propagator $G$ given by 
\begin{equation} 
G_{ij,mn}(t,0) = \int {\cal D}x {\cal D}x^\prime 
{\cal A} [x] {\cal A}^* [x']  {\cal F}_{\rm FV}[x,x^\prime ]  . 
\label{2.2} 
\end{equation} 
The functional ${\cal A} [x]$ denotes the probability amplitude for the 
free system to follow the path $x(t)$ and 
${\cal F}_{FV}[x,x^\prime ]$ denotes the Feynman-Vernon influence functional 
\cite{Feynman63} (see Ref.\  \cite{Weiss99} for details). The functional 
integrations in Eq.\  (\ref{2.2}) extend over paths with endpoints 
$x(0)=x_m, x(t)=x_i, x'(0)=x_n$ and $x'(t)=x_j$ which belong to the 
initial and final states,  $\rho_{mn} (0)$ and $\rho_{ij} (t)$, respectively. 

The technique which we use to calculate the reduced density operator 
Eq.\ (\ref{2.1}) is the iterative tensor multiplication scheme derived 
for the so-called quasiadiabatic propagator path integral (QUAPI). This 
numerical algorithm was developed by Makri and Makarov \cite{quapi} within 
the context of chemical physics. Since its first applications it has been 
succesfully tested and 
adopted to various problems of open quantum systems, with and without 
external driving \cite{quapi,ThorwartQUAPI,ThorwartQUAPI2}. 
Because the details of this 
algorithm has been extensively discussed previously in the literature 
 \cite{quapi,ThorwartQUAPI,ThorwartQUAPI2}, 
 we only mention those prominent ingredients 
 which are of importance for our work:

(i) Symmetric splitting of the short-time propagator: 
To obtain a numerical iteration scheme, we discretize the time interval 
$[0,t]$ into 
${\cal N}$ steps $\Delta t$, such that $t_k =  k \Delta t$ and 
split symmetrically  the full propagator over 
one time step ${\cal U}(t_{k+1},t_k)$ in Eq.\  (\ref{rhored}) according 
to the Trotter formula into a system and an environmental part: 
\begin{eqnarray} 
{\cal U}(t_{k+1},t_k) & \approx & \exp (-i {H}_{\rm B} \Delta t 
/2 \hbar) {\cal U}_{\rm S}(t_{k+1},t_k) \exp (-i {H}_{\rm B} \Delta t
/2 \hbar) \ , \nonumber\\ 
{\cal U}_{\rm S}(t_{k+1},t_k) & = & {\cal T} \exp \left \{ -\frac{i}{\hbar} 
\int_{t_k}^{t_{k+1}} dt' \, 
{H}_{\rm XOR}(t') \right \} \ .
\label{split1} 
\end{eqnarray} 
The neglect of higher order terms of the propagator in Eq.\  (\ref{split1}) 
causes an error of the order of $\Delta t ^3$. 
The short-time propagator ${\cal U}_{\rm S}$ of the bare system 
is given by the corresponding exact system propagators in Eq.\ 
 (\ref{singprop}) over a time step $\Delta t$. 
At this point, we emphasize that this method is not plagued by the 
problem of lacking positivity of the density operator at short times, 
as it is the case for the usually employed master equation approach 
in the Born-Markov limit \cite{Loss98a,Ahn00}. The {\em exact\/} 
coherent dynamics of the bare system enters  and the decomposition of 
the short-time propagator according to Eq.\  (\ref{split1}) is valid for 
any arbitrary short time. 

(ii) The interaction with the bath induces correlations among the paths 
(memory) which are described by the influence functional 
in Eq.\  (\ref{2.2}). As long as the temperature of the Ohmic bath 
is finite, these correlations decay exponentially fast with increasing 
time \cite{Weiss99}. This motivates to neglect 
such long-time correlations and to break up the influence 
kernels into smaller pieces of length $K \, \Delta t$, where 
$K$ denotes the number of time steps over which the memory is fully taken 
into account. 

The two strategies in (i) and (ii) are countercurrent: In step (i) a small 
time step $\Delta t$ is desirable in order to minimizes the error 
due to the neglected 
higher order terms in the propagator. On the other hand, in (ii) 
a large time step is wanted in order to take a long memory range into account. 
A compromise between those two errors has to be found in practice by 
applying the principle of minimal sensitivity \cite{ThorwartQUAPI2} 
to adjust the two parameters $\Delta t$ and $K$, see discussion below.

(iii) The third important ingredient is the appropriate choice of 
basis representation of the problem. For the algorithm it is required 
to iterate the dynamics in the eigenbasis of that system operator which 
couples to the bath. Then the influence functional in Eq.\  (\ref{2.2}) 
can be evaluated in terms of the eigenvalues of the coupling operator. 
In problems 
where the coordinate of a quantum particle in a  continuous potential 
is damped, the continuous position operator turns into a discrete set of 
position eigenvalues. Hence, this representation has been termed 
{\em discrete variable representation (DVR)\/}. \\
{\em Bit-flip errors:\/} If the $\sigma_x$-components of each spin 
couple to the bath, see Eq.\ (\ref{hamint}), the eigenbasis of the 
coupling operator is determined by 
$\mel{\alpha_i}{\left(\sigma_1^{x}+\sigma_2^{x}\right)/2}{\alpha_j} = 
\lambda_i \, \delta_{ij}$ with $\lambda_1 = 0, \lambda_2 = -1,\lambda_3=1$ 
and $\lambda_4 = 0$. A basis rotation of the computational 
basis with basis states $\ket{b_j}$ has to be performed according to 
\begin{equation} 
\ket{\alpha_i} = \sum_{j=1}^4 R_{ij} \ket{b_j}\, ,
\label{basistrans2} 
\end{equation} 
with the transformation matrix 
\begin{equation}
R = \frac{1}{2} \left( 
\begin{array}{rrrr}
1 & 1 & \mbox{\hspace{2ex}} 1 & -1 \\
-1 & -1 & \mbox{\hspace{2ex}} 1 & -1 \\
1 & -1 & \mbox{\hspace{2ex}} 1& 1 \\
-1 & 1 & \mbox{\hspace{2ex}} 1& 1 
\end{array}
\right) \, . 
\label{bastrans}
\end{equation}
\noindent 
{\em Phase errors:\/} For the second case that the $\sigma_z$-components 
of each spin couple to the bath in Eq.\ (\ref{hamint}), the system 
part of  Hamiltonian is already diagonal in the computational basis, i.e.,  
$\mel{b_i}{\left(\sigma_1^{z}+\sigma_2^{z}\right)/2}{b_j} = 
\lambda_i \, \delta_{ij}$ with $\lambda_1 = 1, \lambda_2 = 0,\lambda_3=0$ 
and $\lambda_4 = -1$. No additional basis transformation is necessary. 
\section{Characteristic gate quantifiers}
\label{sec.IV}
In order to quantify the {\em quality\/} of the quantum gate, we use four 
global 
parameters which have been defined by Poyatos, Cirac and Zoller 
\cite{Poyatos97}: (i) the gate fidelity ${\cal F}$, 
(ii) the gate purity ${\cal P}$,(iii) the quantum degree ${\cal Q}$ of 
the gate, and (iv) the entanglement capability ${\cal C}$ of the gate. 
These four quantifiers can be calculated once the reduced density operator 
$\bfrho$ in Eq.\  (\ref{rhored}) is determined. 
To this end, 16 unentangled input states $\ket{\Psi_{\rm in}^j}, j=1,...,16$ 
are defined according to 
$|\psi_a\rangle_1 |\psi_b\rangle_2$ ($a,b=1,...,4$), 
with $\ket{\psi_1}=\ket{0}, \ket{\psi_2}=\ket{1}, 
\ket{\psi_3}=(\ket{0}+\ket{1})/\sqrt{2}$ and 
$\ket{\psi_4}=(\ket{0}+ i \ket{1})/\sqrt{2}\, $. They form one possible 
basis set  and span the Hilbert space for the 
superoperator ${\cal V}_{\rm XOR}$ where 
$\rho(t_{\rm XOR})={\cal V}_{\rm XOR} \, \rho(0)$, see 
Ref.\ \cite{Poyatos97} for details. Moreover, these basis states 
are chosen to be unentangled states in order to avoid the application of a 
preceding two-qubit gate for the preparation of the system state. 

The gate fidelity ${\cal F}$ is defined as the overlapp between the propagation 
with the ideal propagator $U_{\rm XOR}$, Eq.\  (\ref{uxor}), 
averaged over all 16 initial states $\ket{\Psi_{\rm in}^j}$, 
according to
\begin{equation}
{\cal F} = \frac{1}{16}  \sum_{j=1}^{16} 
\mel{\Psi_{\rm in}^j}{U_{\rm XOR}^{+} 
\bfrho_{\rm XOR}^j U_{\rm XOR}}{\Psi_{\rm in}^j} \, 
\label{fidel}
\end{equation}
with $\bfrho_{\rm XOR}^j=\bfrho(t_{\rm XOR})$ with initial condition 
$\bfrho(0)= \ket{\Psi_{\rm in}^j}\bra{\Psi_{\rm in}^j}$. 

In a similar way, the purity ${\cal P}$ is defined as 
\begin{equation}
{\cal P} = \frac{1}{16}  \sum_{j=1}^{16} \tr (\bfrho_{\rm XOR}^j)^2
 \, .
\label{pur}
\end{equation}
This quantity is proportional to the (negative) linearized entropy and 
reflects the effects of decoherence. 

The third quantity, the quantum degree ${\cal Q}$ of the gate, is defined 
as the maximum of the overlap of all possible output states stemming from 
 unentangled states and of all maximally entangled (Bell) states 
$\ket{\Psi_{\rm me}^k}, k=1,...,4$. In formal 
 terms, this implies 
\begin{equation}
{\cal Q} = \max_{j, k} \mel{\Psi_{\rm me}^k}{\bfrho_{\rm XOR}^j}{\Psi_{\rm me}^k}
 \, .
\label{qdeg}
\end{equation}
The purpose of this parameter is to quantify the notion of nonlocality. 
Bennett and co-workers \cite{Bennett96} have shown that all those density 
operators  which have an overlap with a maximally entangled state 
being larger than the value $(2+3 \sqrt{2})/8 \approx 0.78$ are non-local, i.e., 
violate the Clauser-Horne-Shimony-Holt inequality \cite{Bennett96}. 

Obviously, ${\cal F} =1, {\cal P} =1,{\cal Q} =1$ denote the ideal 
gate operation.

The fourth quantifier is the entanglement capability ${\cal C}$ of the gate. 
It denotes the smallest eigenvalue of the partial transposed density 
matrix \cite{Horo96} which is determined from $\bfrho_{\rm XOR}^j$ for all 
unentangled input states $\ket{\Psi_{\rm in}^j}$. $\bfrho_{\rm XOR}^j$ 
characterizes an entangled state if and only if the smallest 
eigenvalue of the partial transposed density operator is negative. 
The ideal operation has an entanglement capability of ${\cal C}=-0.5$. 
\section{Results}
\label{sec.V}
Having determined the reduced matrix in Eq.\  (\ref{rhored}) by the iterative 
QUAPI algorithm, we investigate the influence of the interaction 
with the environment systematically. Therefore, we assume that the two qubits 
are identical and experience external fields of the same strength, i.e., 
$B_1^x=B_2^x=B^x$ and $B_1^z=B_2^z=B^z$. Moreover, we introduce the 
following dimensionless parameters: We
  scale the quantities with respect to the characteristic 
 energy scale of the single qubit which is given by the  
 energy splitting $\hbar B^z$ of the single qubit. This in turn defines 
 a time scale $(B^z)^{-1}$. Consequently, the temperature is given 
 in units of $\hbar B^z/k_{\rm B}$ (Note that $\gamma^{x/z}$ is already 
 dimensionless). For all following results, we have used 
 a cut-off frequency of $\omega_c =50 B^z$ in (\ref{contspede}).  

\subsection{Time resolved quantum XOR operation}
\label{sec.Va}
\begin{figure}[th]
\begin{center}
\epsfig{figure=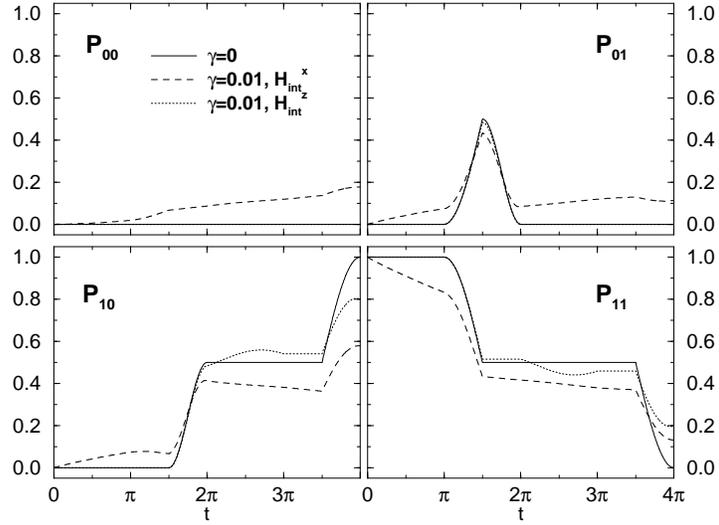,width=90mm,keepaspectratio="true"} 
\end{center}
\caption{Time resolved dynamics of the quantum XOR operation for the 
case of equal energies, i.e.,  $B^x=B^z=J$. 
Depicted are the populations $P_{ij}(t) = \mel{ij}{\bfrho(t)}{ij}$ as 
a function 
of time for the initial condition $\bfrho(0)=\ket{11}\bra{11}$ for 
three different cases of (i) no damping, $\gamma^{x/z}=0$, (solid line), 
(ii) bit flip errors with $\gamma^x=0.01$ (long dashed line), 
and (iii) phase errors with $\gamma^z=0.01$ (dotted line). 
The time is scaled in units of $(B^z)^{-1}$. 
Moreover, we set the temperature 
to $T=0.1 \hbar B^z/k_{\rm B}$ and the cut-off frequency to $\omega_c 
= 50 B^z$.  \label{fig.xor}}
\end{figure}

We first illustrate the 
time-resolved dynamics of a generic XOR operation. 
To this end, we determine the populations of the four states of the 
computational basis as a function of time, i.e., 
$P_{ij}(t) := \mel{ij}{\bfrho(t)}{ij}$ with $i,j=0,1$ for the initial 
condition $\bfrho(0)=\ket{11}\bra{11}$. We choose the pulse sequence 
sketched in Fig.\  \ref{fig.pulse} with 
parameters set to $B^x=B^z$ and $J=B^z$. 
Moreover, we choose for illustrative purpose a rather high temperature 
of $T=0.1\hbar B^z/k_{\rm B}$. 
Fig.\  \ref{fig.xor} depicts the 
results for the three different cases of 
(i) no damping, $\gamma^{x/z}=0$, (solid line), 
(ii) bit flip errors with $\gamma^{x}=0.01$ (long dashed line), 
and (iii) phase errors with $\gamma^{z}=0.01$ (dotted line).  
The switching times $t_j$  are equal to multiples of $\pi/2$ 
for  this special case of equal energies. 

The iterative QUAPI algorithm possesses two 
free parameters which have to be properly adjusted. We fix the 
number $K$ of memory time steps and the length $\Delta t$ of each time 
step according to the {\em principle of minimal sensitivity\/} 
\cite{ThorwartQUAPI2}. By applying this method, we obtain the values 
 $\Delta t=0.15 (B^z)^{-1}$ with $K=2$ (not shown). 

As one observes, 
the final state of the ideal operation ($\gamma^{x/z}=0$) 
is $\ket{\Psi}=\ket{10}$.
The deviation of the dynamics in presence of decoherence and dissipation 
from the ideal case is clearly visible. 
\subsection{Quality of realistic quantum XOR operations}
\label{sec.Vb}
In order to study realistic physical situations, we apply our method to 
three different condensed matter systems, namely  flux qubits 
(set I) \cite{Mooij99,Makhlin00}, 
charge qubits (set II)  
\cite{Makhlin00,Averin98}, 
and qubits realized in  coupled semiconductor quantum 
dots (set III)  
\cite{Loss98a,Loss99a}. 
We choose typical parameter sets published in the literature. They 
 are summarized in Table \ref{tab.sets}. 
 
 The QUAPI parameters are again determined by the 
 principle of minimal sensitivity for $K=3$. We obtain for the 
 case of $H_{\rm int} ^{x}$ (bit flip errors) for set I 
 (in units of $(B^z)^{-1}$): $\Delta t =0.06$, 
 for set II: $\Delta t =0.2$, and for set III: $\Delta t =0.02$, 
 and for the 
 case of $H_{\rm int} ^{z}$ (phase errors) for set I: $\Delta t =0.013$, 
 for set II: $\Delta t =0.08$, and for set III: $\Delta t =0.01$. 
\subsubsection{Dependence on temperature}
\begin{figure}[th]
\begin{center}
\psfrag{FF}{$\cal{F}$}
\psfrag{PP}{$\cal{P}$}
\psfrag{QQ}{$\cal{Q}$}
\psfrag{CC}{$\cal{C}$}
\epsfig{figure=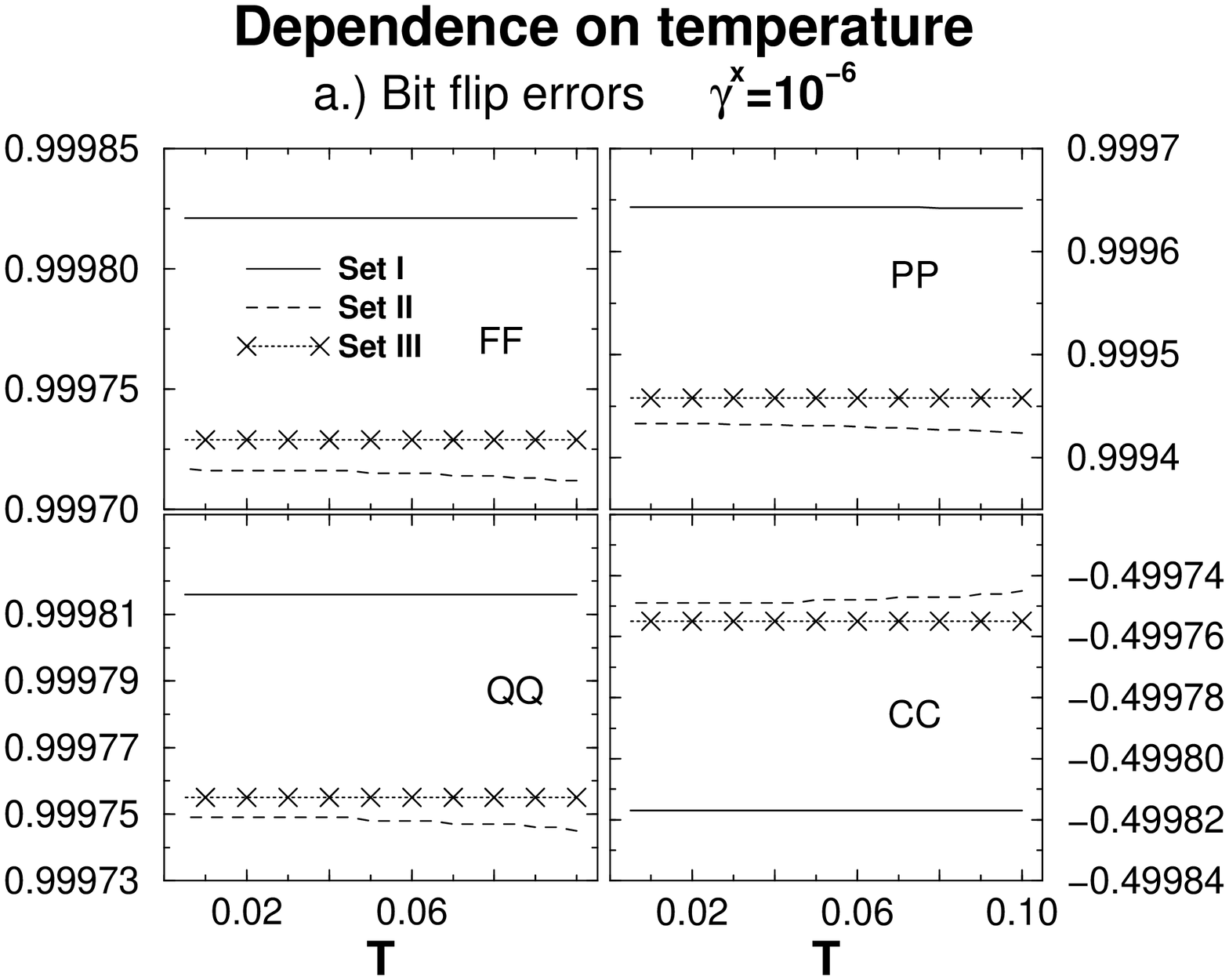,width=90mm,keepaspectratio="true"} 
\epsfig{figure=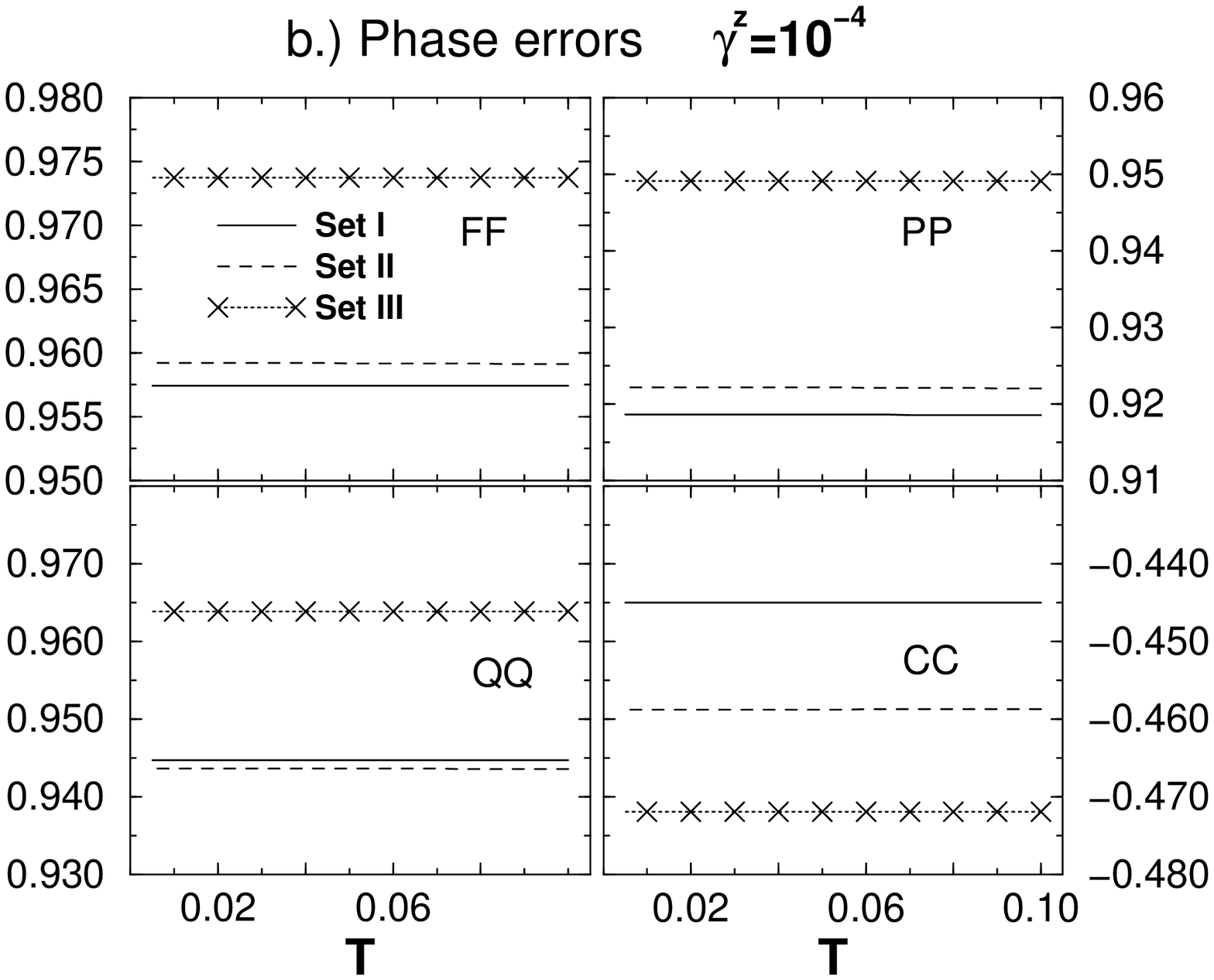,width=90mm,keepaspectratio="true"} 
\end{center}
\caption{Dependence of the fidelty ${\cal F}$, the purity ${\cal P}$, 
the quantum degree ${\cal Q}$, see (\ref{fidel} - \ref{qdeg}) 
and the entanglement capability 
${\cal C}$ on temperature $T$ for bit flip errors (Fig.\ \ref{fig.tempx} a.) 
and phase errors (Fig.\ \ref{fig.tempx} b.). The temperature is 
scaled in units of $\hbar B^z / k_{\rm B}$. The qubit parameters 
are given in Table \ref{tab.sets}.  The damping constant for the 
bit flip errors is set at $\gamma^{x}=10^{-6}$ and for the phase errors at 
$\gamma^{z}=10^{-4}$. 
 \label{fig.tempx}}
\end{figure}

The dependence of the four characteristic gate quantifiers on the 
bath temperature $T$ is depicted with Fig.\  \ref{fig.tempx}. 
In panel a.) the 
influence of the random bit flips are investigated while the panel 
b.) depicts the effect of phase errors. The damping constant for the 
bit flip errors is chosen to be $\gamma^{x}=10^{-6}$ and for the phase errors 
 we seet $\gamma^{z}=10^{-4}$ \cite{Makhlin00}. 

First, one observes that all results depend only weakly on temperature. 
Extrapolating the results to zero temperature indicates the influence 
of the zero point oscillations of the harmonic bath oscillators. Second, 
the results for ${\cal F},{\cal P}$ and ${\cal Q}$ do in no case exceed a 
value of 0.999 85 for bit flip errors and 0.975 for 
phase errors. This demonstrates that even smaller strengths of the 
coupling to the environment than $\gamma^{x}=10^{-6}$ or $\gamma^{z}=10^{-4}$ 
are necessary in order to obtain a desired value of 0.999 99 \cite{Loss98a}! 

Furthermore, we observe that the flux qubits (set I) are the least sensitive 
to bit flip errors while charge qubits (set II) and quantum dot qubits 
(set III) both are performing worse, see Fig.\  
\ref{fig.tempx} a.). On the other hand, 
quantum dot qubits (set III) are most immune from 
phase errors, see Fig.\  \ref{fig.tempx} b.) while the superconducting systems 
assume worse results. 
\subsubsection{Dependence on the damping strength}
\begin{figure}[th]
\begin{center}
\psfrag{FF}{$\cal{F}$}
\psfrag{PP}{$\cal{P}$}
\psfrag{QQ}{$\cal{Q}$}
\psfrag{CC}{$\cal{C}$}
\epsfig{figure=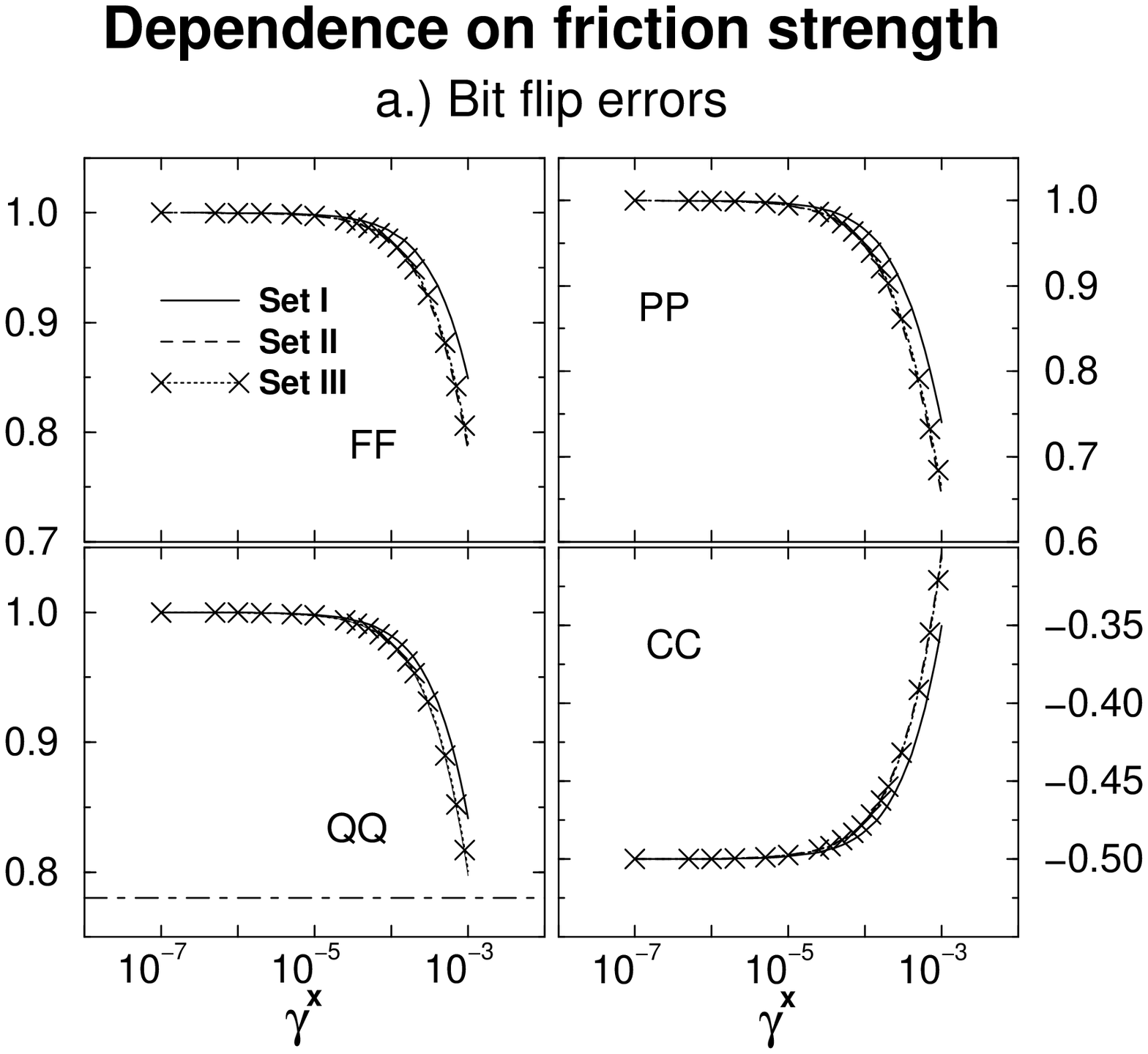,width=90mm,keepaspectratio="true"} 
\epsfig{figure=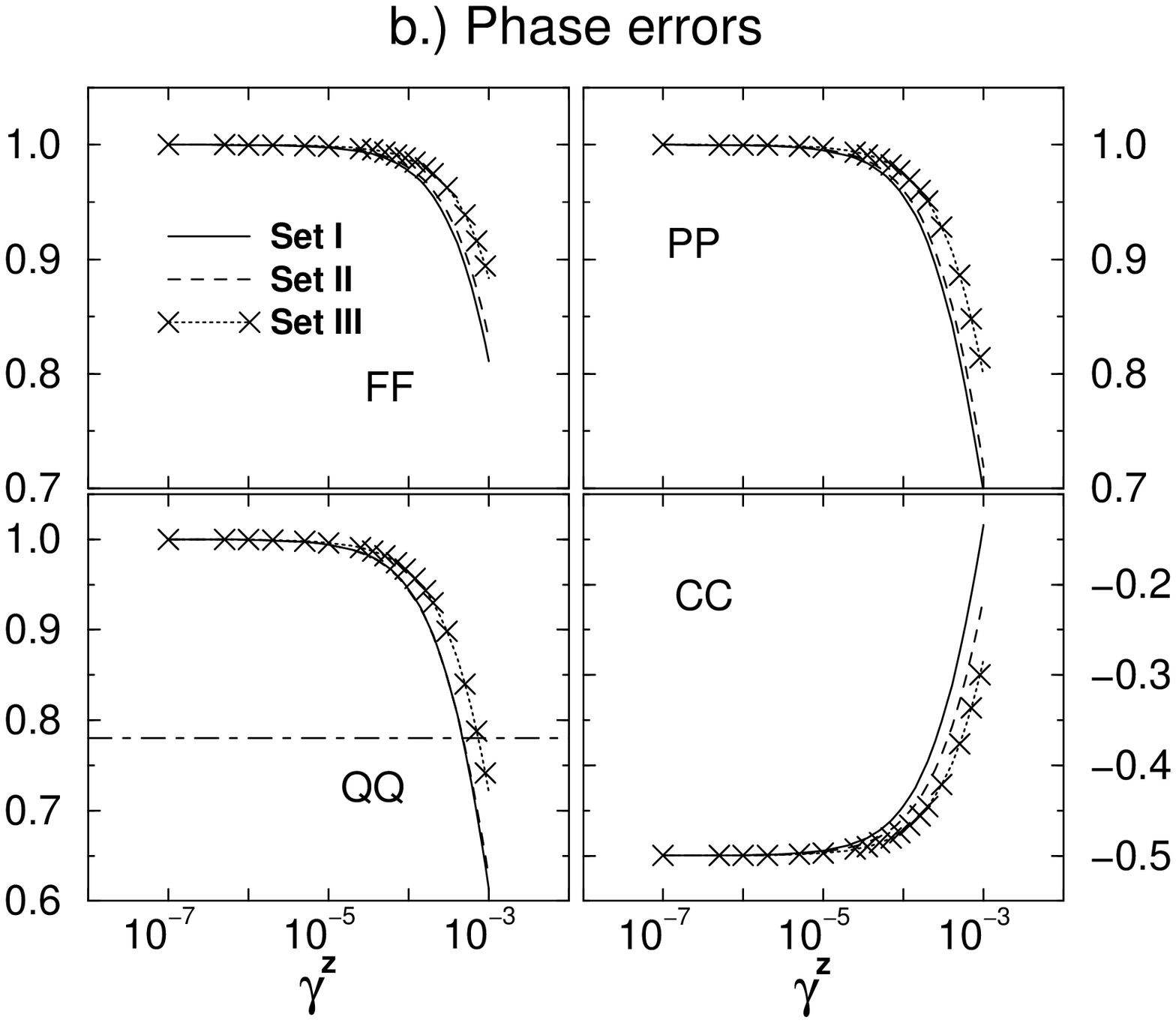,width=90mm,keepaspectratio="true"} 
\end{center}
\caption{Dependence of the four gate quantifiers on the 
dimensionless friction strength. Upper panel a.): 
Bit flip errors ($\gamma^x$), lower panel b.): 
Phase errors ($\gamma^z$). The lower bound of 
${\cal Q}\approx 0.78$ for the Clauser-Horne-Shimony-Holt inequality is 
indicated by the horizontal dotted-dashed line (see text). 
For the remaining parameters, see 
Table \ref{tab.sets}. \label{fig.gamx}}
\end{figure}

As shown in the preceding section, the quality of the gate 
operation cannot be improved by lowering the temperature of the environment. 
On the contrary, the shielding of the qubit system against external noise  
is most important. This is demonstrated by the results for the 
dependence of the four gate quantifiers on the coupling 
constants $\gamma^{x/z}$ depicted in Fig.\  \ref{fig.gamx}. 

We find again that the results for the flux qubits (set I) approach 
the ideal values of ${\cal F}={\cal P}={\cal Q}=1$ and 
${\cal C}=-0.5$ first upon decreasing the friction strength, 
see Fig.\  \ref{fig.gamx} a.). 
However, the flux qubits are more sensitive to  phase errors from 
which the quantum dot qubits (set III) are seemingly most immune, see Fig.\  
\ref{fig.gamx} b.). Note that in Fig.\ \ref{fig.gamx} the lower bound of 
${\cal Q}\approx 0.78$ for the Clauser-Horne-Shimony-Holt inequality is 
indicated by the horizontal dotted-dashed line. 

For all three systems we find that the damping strength  
has to be less than $\gamma^{x/z}\approx 10^{-7}$ in order to serve as  
 suitable candidates for a quantum information processor. 
\subsubsection{Dependence on interqubit coupling strength}
\begin{figure}[th]
\begin{center}
\psfrag{FF}{$\cal{F}$}
\psfrag{PP}{$\cal{P}$}
\psfrag{QQ}{$\cal{Q}$}
\psfrag{CC}{$\cal{C}$}
\epsfig{figure=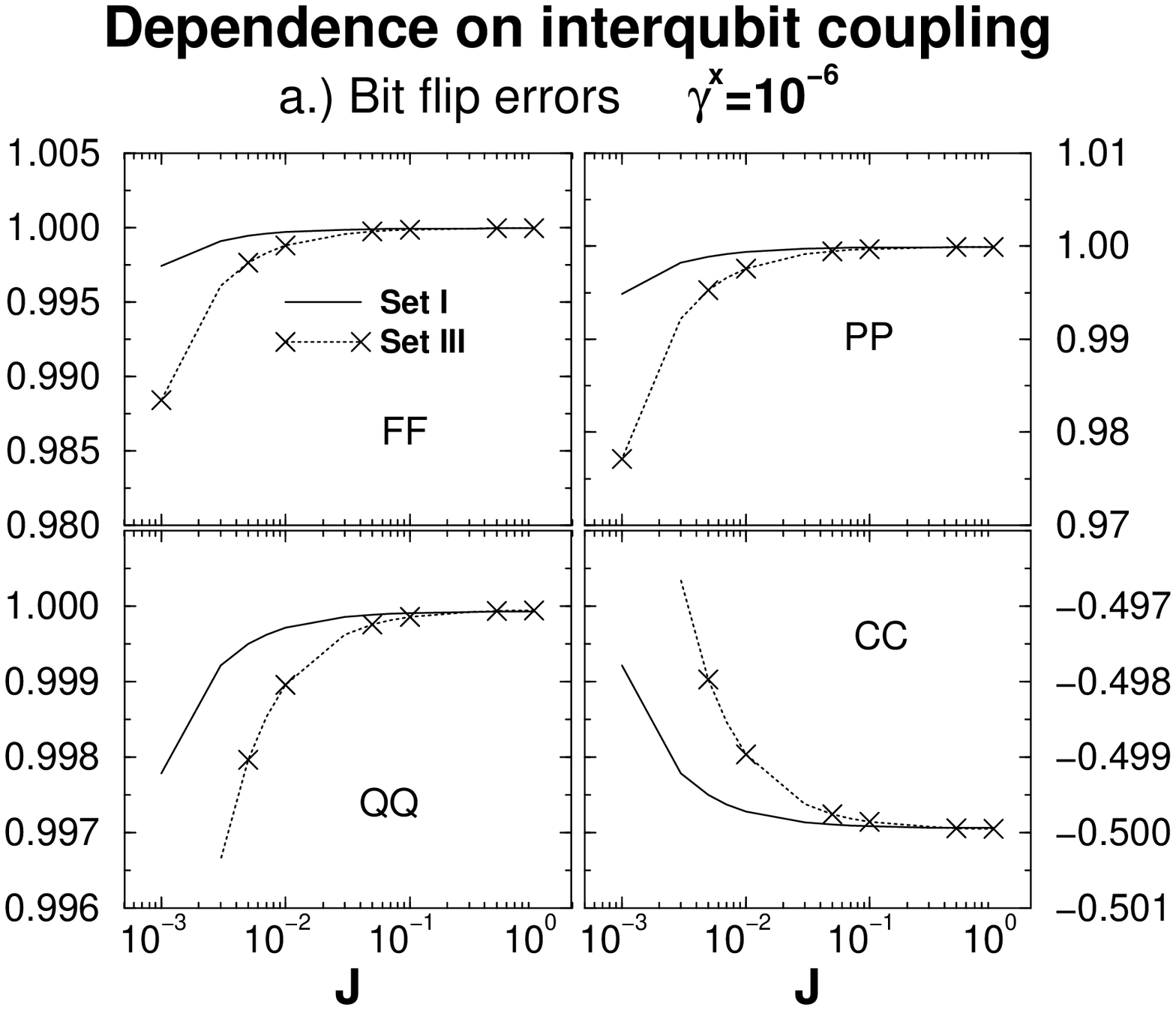,width=90mm,keepaspectratio="true"} 
\epsfig{figure=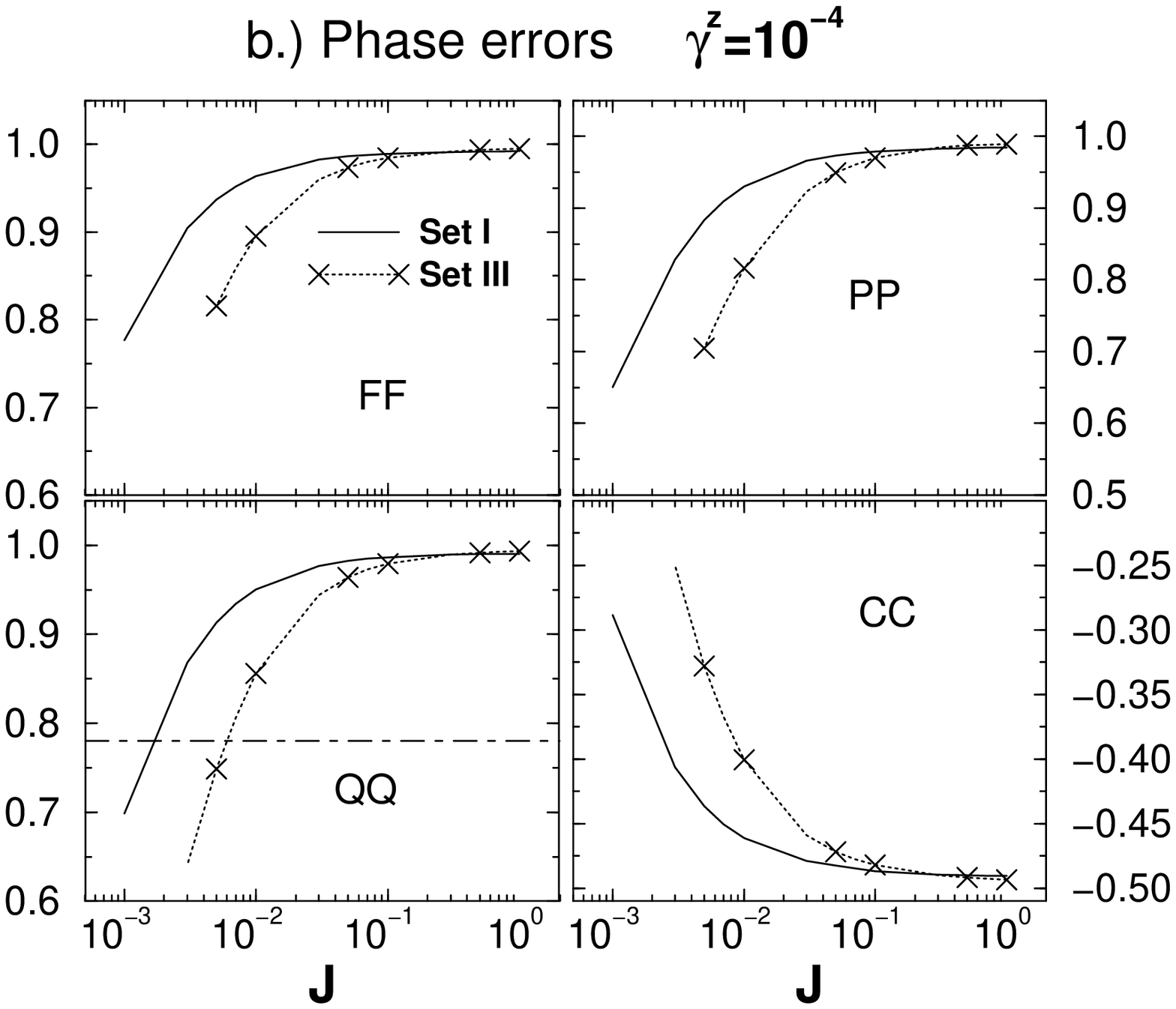,width=90mm,keepaspectratio="true"} 
\end{center}
\caption{Dependence of the gate quantifiers on the 
strength $J$ of the interqubit coupling. Depicted are the results for 
parameter set I and III from Table  \ref{tab.sets}. The upper 
panel a.) shows the results for the bit-flip error 
with $\gamma^x=10^{-6}$ while the lower panel b.) depicts the results 
for the phase error with $\gamma^z=10^{-4}$. The horizontal dotted-dashed line 
marks the lower bound of 
${\cal Q}\approx 0.78$ for the Clauser-Horne-Shimony-Holt inequality. 
\label{fig.couplx}}
\end{figure}

In the remaining part we address the dependence of the quality of the gate 
operation on the strength $J$ of the interqubit coupling. 
Physically, one expects that a coupling strength which is comparable to 
the characteristic qubit energy ($J\approx B^z$) would yield the best results. 
This is confirmed in Fig.\  \ref{fig.couplx} for the two 
different sets I and III (set II is equivalent to set I). The 
panel a.) illustrates the results for the bit flip errors 
with $\gamma^x=10^{-6}$,  and correspondingly 
 panel b.) for the phase errors with 
$\gamma^z=10^{-4}$. As one can observe, the superconducting qubits (set I) can 
perform with a smaller interqubit coupling than the quantum dot qubits 
(set III). Moreover, the use of an interqubit coupling larger than  
$J\approx 10^{-1}B^z$ is advantageous. 
The horizontal dotted-dashed line in Fig.\ \ref{fig.couplx} b.) indicates 
the lower bound of ${\cal Q}\approx 0.78$ for nonlocality.
\section{Conclusions}
\label{sec.VI}
In this work we have shown that the numerical quasiadiabatic 
propagator path integral method (QUAPI) of Makri and Makarov 
provides an appropriate method to investigate decohering 
quantum information processes 
which involve time-dependent Hamiltonians in presence of a coupling to 
an external environment. We have applied this iterative algorithm to the 
example of the quantum XOR gate operation and have obtained the 
full time-resolved evolution of the two-qubit system in presence of 
time-dependent external fields. No further approximations on the time evolution 
of the gate operation such as a Markovian evolution or extended time 
spans of the gate operation have been invoked.  

We have investigated the quality of the quantum XOR operation 
by numerically 
determining four characteristic gate quantifiers, i.e., the fidelity 
${\cal F}$, the purity ${\cal P}$, the quantum degree ${\cal Q}$, and 
the entanglement capability  ${\cal C}$ of the gate. We have simulated 
the gate operation for three parameter sets which mimic  
realistic experimental situations for condensed matter qubits, such as  
flux and charge qubits in superconducting Josephson systems and 
qubits in semiconductor quantum dots. Thereby, we have suceeded to 
investigate  
systematically  the quality of the gate operation as a function of 
the environmental temperature $T$, 
 the friction strength $\gamma$, 
and the strength $J$ of the interqubit coupling. Two different types 
of errors in the qubits have been modelled: (i) bit-flip errors 
and (ii) phase errors. We have elucidated 
 how the different physical setups perform 
under these conditions. 
As major findings we establish that the quality of the gate depends 
{\em only weakly on temperature\/} but rather {\em strongly on the 
friction strength\/}. 
Moreover, we have illustrated that the interqubit coupling 
strength plays an important role and should not be smaller than $0.1E_0$ with  
$E_0$ being the typical energy scale of the qubit system.  


In order for a quantum information processor to operate optimally 
 the decoherent 
influence of the environment needs to be suppressed. Therefore, three different 
approaches are currently discussed: These are the techniques of quantum 
error correction, fault tolerant quantum computation, 
and entanglement purification \cite{qivBuch}. 
The general idea common to all three methods is to use 
for quantum information processing only a small subset 
of a larger set of entangled ancilla qubits.  Although these ideas are 
very promising for small register lengths, the techniques become 
increasingly difficult if one attempts to realize large qubit registers 
in physical systems. Moreover, one has to keep track of the quantum 
state of the environment. Whilst these requirements are seemingly feasible 
for quantum optical information processing 
systems \cite{Wineland00}, they appear insufficient for  
condensed matter systems with their characteristic huge number of 
environmental degrees of freedom. 

An alternative approach consists in minimizing the occurrence of errors by 
{\em controlling decoherence\/} via the application of tailored 
 time-dependent external fields 
to qubit systems \cite{Grifoni98,Oelschlaegel93,Hanggi95,Viola99,jmodopt00}. 
For example, the application of a time-dependent periodic external field can 
induce a Floquet spectrum with degenerate quasienergy states 
\cite{Grifoni98,Oelschlaegel93,Hanggi95} or it can move the qubit out off 
resonance with certain bath modes \cite{jmodopt00} thereby reducing 
decoherence. The suitability of such schemes 
 to a quantum gate operation, however, remains 
to be demonstrated. 
\section*{Acknowledgement}
We thank Yu.\ Makhlin and R.\ Blick for helpful discussions. 
This work has been supported by the Deutsche Forschungsgemeinschaft via Grant 
No.\ HA 1517/19-1.

\newpage

\begin{table}[th]
\begin{tabular}{cccccccc}
Set & $B^z$ & $B^x$ & $J$ & $T$ & $B^x/B^z$ & $J/B^z$ & $T/B^z$ \\
\hline
I: Flux qubits & 0.5 K & 50 mK & 25 mK & 25 mK & 0.1 & 0.05 & 0.05 \\
II: Charge qubits & 1 K & 100 mK & 5 mK & 50 mK & 0.1 & 0.005 & 0.05 \\
III: Quantum dot qubits & 
1 meV & 1 meV & 0.05 meV & 125 mK & 1 & 0.05 & 0.01 \\
\end{tabular}
\caption{Parameter sets which have been used for the simulations of the 
quantum XOR gate. They mimic typical experimental situations in 
three important solid-state qubit systems, i.e., flux qubits and charge qubits 
in superconducting Josephson devices (Set I and II) and 
spin and charge qubits in ultrasmall semiconductor quantum dots (Set III). 
 \label{tab.sets}}
\end{table}

\newpage

\newpage

\end{document}